# Superconducting Surface Impedance under Radiofrequency Field

B. P. Xiao,[1,2] C. E. Reece,[1,*] and M. J. Kelley[1,2]

[1]*Thomas Jefferson National Accelerator Facility, Newport News, VA 23606, U.S.A.*
[2]*College of William and Mary, Williamsburg, VA 23187, U.S.A.*

Based on BCS theory with moving Cooper pairs, the electron states distribution at 0K and the probability of electron occupation with finite temperature have been derived and applied to anomalous skin effect theory to obtain the surface impedance of a superconductor under radiofrequency (RF) field. We present the numerical results for Nb and compare these with representative RF field-dependent effective surface resistance measurements from a 1.5 GHz resonant structure.

## Introduction

The radiofrequency (RF) surface impedance of a superconductor may be considered a consequence of the inertia of the Cooper pairs in the superconductor. The resulting incomplete shielding of RF field allows the superconductor to store RF energy inside its surface, which may be represented by surface reactance. The RF field that enters the superconductor will interact with quasi-particles, causing RF power dissipation, represented by surface resistance. Based on the BCS theory [1] and anomalous skin effect theory [2], a derivation of a superconductor's surface impedance was developed by Mattis and Bardeen [2, 3]. In reference [4], the Mattis-Bardeen theory has been written in terms of the Fourier components of current $\boldsymbol{j(p)}$ and vector potential $\boldsymbol{A(p)}$ by defining $K(p)$ as,

$$\boldsymbol{j(p)} = -\frac{c}{4\pi} K(p) \boldsymbol{A(p)} \tag{1}$$

where

$$K(p) = \frac{-3}{4\pi \hbar v_F \lambda_L^2(0)} \int_0^\infty \int_{-1}^1 e^{ipRu} e^{-\frac{R}{l}} (1-u^2) \times I(\omega, R, T) du dR = \frac{-3}{4\pi \hbar v_F \lambda_L^2(0)} \int_0^\infty \frac{4}{(pR)^2} \left[ \frac{\sin(pR)}{pR} - \cos(pR) \right] e^{-\frac{R}{l}} \times I(\omega, R, T) dR \tag{2}$$

with $v_F$ the Fermi velocity, $\lambda_L(0)$ the London penetration depth at 0 K, and $\omega$ the angular frequency of the RF field. The integrations over $R$ and $u$ are the space and angular integrations, respectively. The term $I(\omega,R,T)$ is calculated from the single-particle scattering operator based on BCS theory.

The surface impedance of a superconductor with random scattering at the surface is [4],

$$Z = i\pi\omega\mu_0 \left\{ \int_0^\infty \ln[1 + K(p)/p^2] dp \right\}^{-1} \tag{3}$$

Mattis-Bardeen theory, however, does not consider the field dependence of surface impedance. In particular, its real part, surface resistance, which is of great interest in superconducting radiofrequency (SRF) applications, is unaddressed. Several models have been proposed to address this issue [5, 6, 7]. Preliminary work on the field dependence of Mattis-Bardeen theory has been conducted based on Rogers and Bardeen's work, the field dependence of surface resistance has been calculated with simplified models: Kulik and Palmieri [6] use a model in which the energy gap could be reduced by the existence of an RF field, $\Delta = \Delta_0 - p_F v_S(H)$ [8], with $\Delta$ the energy gap with field, $\Delta_0$ the energy gap without field, $p_F$ the Fermi momentum and $v_s$ the Cooper pair velocity; and Gurevich [7] used a model based on the field dependence of the quasi-particle distribution function [8].

Here, starting from the BCS theory with a net current in a superconductor, we calculate the electron states distribution at 0 K and the probability of electron occupation with finite temperature and applied to anomalous skin effect theory, which describes the response of metals to high frequency electromagnetic field at low temperature, to obtain a new form of RF field dependence of the surface impedance.

## Electron states distribution with net current flow

In BCS theory, paired particles in the ground state, with total mass $2m$ and zero total momentum that occupy state ($k \uparrow$, $-k \downarrow$), with velocity $v_k$ in random direction, and energy relative to the Fermi sea $\varepsilon_F$ of $\varepsilon_k$, have been considered to give minimum free energy for superconductors. States with a net flow in a certain direction can be obtained by taking a pairing ($k+q \uparrow$, $-k+q \downarrow$), with total momentum $2q$ the same for all Cooper pairs, corresponding to net velocity $v_s = \hbar q/m$.

With temperature $T$ close to 0 and a net current flow, the probability of the state ($k+q \uparrow$, $-k+q \downarrow$) being occupied by a pair of particles is $h_k$. With a finite $T$, single electrons start to appear. $f_{-k+q \downarrow}$ is defined as the probability that state $-k+q \downarrow$, with Bloch energy relative to the Fermi sea of $\varepsilon_{-k+q}$, being occupied, and $f_{k+q \uparrow}$, that probability for state $k+q \uparrow$, with Bloch energy relative to the Fermi sea of $\varepsilon_{k+q}$. We have then,

$$\varepsilon_{k+q} = \frac{1}{2}m(v_k + v_s)^2 - \varepsilon_F = \varepsilon_k + \varepsilon_s + \varepsilon_{ext} \tag{4}$$

$$\varepsilon_{-k+q} = \frac{1}{2}m(-v_k + v_s)^2 - \varepsilon_F = \varepsilon_k + \varepsilon_s - \varepsilon_{ext} \tag{5}$$

with $\varepsilon_s = \frac{1}{2}m v_s^2$, $\varepsilon_{ext} = m v_k v_s = \sqrt{2m(\varepsilon_F + \varepsilon_k)} v_s x \approx p_F v_s x$ and $x = \cos\alpha$, $\alpha$ is the angle between $v_k$ and $v_s$, from 0 to $\pi$. $k_F$ is Fermi wave vector.

We define $h_k$ the probability that state ($-k+q \downarrow$, $k+q \uparrow$) is occupied by a ground pair while $T \to 0$, and define $s_{-k+q \downarrow}$ the probability that $-k+q \downarrow$ is occupied and $k+q \uparrow$ is empty, $s_{k+q \uparrow}$ for vice versa, $p_k$ the probability that state ($-k+q \downarrow$, $k+q \uparrow$) is occupied by a pair of excited particles, and

$$s_{-k+q \downarrow} = f_{-k+q \downarrow}(1 - f_{k+q \uparrow}) \tag{6}$$

$$s_{k+q \uparrow} = f_{k+q \uparrow}(1 - f_{-k+q \downarrow}) \tag{7}$$

$$p_k = f_{-k+q \downarrow} f_{k+q \uparrow} \tag{8}$$

$$1 - s_{-k+q \downarrow} - s_{k+q \uparrow} - p_k = (1 - f_{-k+q \downarrow})(1 - f_{k+q \uparrow}) \tag{9}$$

$f$ is the distribution function for excited particles. Above the Fermi surface with $k > k_F$ $f$ refers to electron occupation, and below the Fermi surface with $k < k_F$ $f$ refers to hole occupation. (9) is the probability that state ($-k+q \downarrow$, $k+q \uparrow$) is occupied by a pair of ground state particles. For $-k+q \downarrow$ with $k > k_F$, the occupation of single electrons, excited pairs of electrons and ground state pairs of electrons are $s_{-k+q \downarrow}$, $p_k(1-h_k)$ and $(1-f_{-k+q \downarrow})(1-f_{k+q \uparrow})h_k$, respectively. For $-k+q \downarrow$ with $k < k_F$, the occupation of single holes, excited pairs of holes and ground state pairs of holes are $s_{-k+q \downarrow}$, $p_k h_k$ and $(1-f_{-k+q \downarrow})(1-f_{k+q \uparrow})(1-h_k)$, respectively. Similar to (3.16) in reference [1], the total free energy of a superconductor in the superconducting state can then be expressed as:

$$F_s = \sum_{k>k_F} \varepsilon_{-k+q}[s_{-k+q \downarrow} + p_k(1 - h_k) + (1 - f_{-k+q \downarrow})(1 - f_{k+q \uparrow})h_k] + \sum_{k<k_F}(-\varepsilon_{-k+q})[s_{-k+q \downarrow} + p_k h_k + (1 - f_{-k+q \downarrow})(1 - f_{k+q \uparrow})(1 - h_k)] + \sum_{k>k_F} \varepsilon_{k+q}[s_{k+q \uparrow} + p_k(1 - h_k) + (1 - f_{-k+q})(1 - f_{k+q \uparrow})h_k] + \sum_{k<k_F}(-\varepsilon_{k+q})[s_{k+q \uparrow} + p_k h_k + (1 - f_{-k+q})(1 - f_{k+q \uparrow})(1 - h_k)] - \sum_{k,k'} V[h_k(1 - h_k)h_{k'}(1 - h_{k'})]^{1/2}(1 - f_{-k+q \downarrow} - f_{k+q \uparrow})(1 - f_{-k'+q \downarrow} - f_{k'+q \uparrow}) + kT \sum_k [f_{-k+q \downarrow} \ln f_{-k+q \downarrow} + (1 - f_{-k+q \downarrow})\ln(1 - f_{-k+q \downarrow}) + f_{k+q \uparrow} \ln f_{k+q \uparrow} + (1 - f_{k+q \uparrow})\ln(1 - f_{k+q \uparrow})] \tag{10}$$

Applying (6)(7)(8)(9) into (10), by minimizing $F_s$ with respect to $h_k$ and $f_{-k+q\downarrow}$ separately, we obtain:

$$\frac{\sqrt{h_k(1-h_k)}}{1-2h_k} = \frac{\sum_{k'} V[h_{k'}(1-h_{k'})]^{1/2}(1-f_{-k'+q\downarrow}-f_{k'+q\uparrow})}{\varepsilon_{-k+q}+\varepsilon_{k+q}} \qquad (11)$$

$$h_k = \frac{1}{2}\left(1-\frac{\varepsilon_k+\varepsilon_s}{E_k}\right) \qquad (12)$$

$$[h_k(1-h_k)]^{1/2} = \frac{\Delta}{2E_k} \qquad (13)$$

$$f_{-k+q\downarrow} = \begin{cases} f(E_{-k+q\downarrow}), & k > k_F \\ f(E_{k+q\uparrow}), & k < k_F \end{cases} \qquad (14)$$

$$f_{k+q\uparrow} = \begin{cases} f(E_{k+q\uparrow}), & k > k_F \\ f(E_{-k+q\downarrow}), & k < k_F \end{cases} \qquad (15)$$

with $\beta = \frac{1}{kT}$ and

$$\Delta = \sum_{k'} V[h_{k'}(1-h_{k'})]^{1/2}\left(1-f_{-k'+q\downarrow}-f_{k'+q\uparrow}\right) \qquad (16)$$

$$E_{k+q\uparrow} = E_k + \varepsilon_{ext}, \quad E_{-k+q\downarrow} = E_k - \varepsilon_{ext} \qquad (17)$$

$$f(E) = \frac{1}{e^{\beta E}+1} \qquad (18)$$

and $E_k = [(\varepsilon_k+\varepsilon_s)^2+\Delta^2]^{\frac{1}{2}}$ . $\qquad (19)$

The excited particle distribution functions $f_{-k+q\downarrow}$ and $f_{k+q\uparrow}$ are not continuous at $k = k_F$ since they specify electron occupations above Fermi surface and hole occupations below it. The probability that $-k+q\downarrow$ is occupied by an electron and $k+q\uparrow$ is empty, is continuous at $k = k_F$, with a value at $f(E_{-k+q\downarrow})[1-f(E_{k+q\uparrow})]$.

These results are similar to the results in reference [7] using a different approach, with the expression of $E_k$ slightly different.

The functions that describe the probability of quasi-particle (single or excited pair of electrons/holes) distribution and the modified density of states are,

$$\frac{1}{2}\int_{-1}^{1} f_{-k+q\downarrow} dx = \frac{1}{2}\int_{-1}^{1} f_{k+q\uparrow} dx = 1 + \frac{1}{2\beta p_F v_s}\ln\left(\frac{1+e^{\beta(E_k-p_F v_s)}}{1+e^{\beta(E_k+p_F v_s)}}\right) \qquad (20)$$

$$\frac{1}{2}\int_{-1}^{1}\frac{d\varepsilon}{dE_{k+q\uparrow}} dx = \frac{1}{2}\int_{-1}^{1}\frac{d\varepsilon}{dE_{-k+q\downarrow}} dx = \begin{cases} \frac{\sqrt{(E_a+p_F v_s)^2-\Delta^2}-\sqrt{(E_a-p_F v_s)^2-\Delta^2}}{2p_F v_s}, & E_a > \Delta+p_F v_s \\ \frac{\sqrt{(E_a+p_F v_s)^2-\Delta^2}}{2p_F v_s}, & \Delta+p_F v_s > E_a > \Delta-p_F v_s \\ 0, & E_a < \Delta-p_F v_s \end{cases} \qquad (21)$$

Equation (21), with $E_a = E_{k+q\uparrow}$ or $E_a = E_{-k+q\downarrow}$, is the same as the corresponding equation in reference [6]. Equation (19) shows $2\Delta$ energy, which represents the energy gap of this superconducting system, is still necessary for a photon to break a Cooper pair, while quasi-particles can exist within the $2\Delta$ regime nearby the Fermi surface. The minimum gap in energy spectrum for quasi-particles in this superconductor reduces to $2(\Delta-p_F v_s)$, as shown in equations (20) and (21).

The energy gap, as a function of Cooper pair velocity, may be derived by applying (12)(13)(14)(15)(17)(18)(19) into (16) and changing the sum into integration,

$$\frac{1}{NV} = \int_{-1}^{1}\frac{dx}{2}\int_{-\hbar\omega_D}^{\hbar\omega_D}\frac{d\varepsilon_k}{2[(\varepsilon_k+\varepsilon_s)^2+\Delta^2]^{1/2}}\left(1-\frac{1}{e^{\beta\left\{[(\varepsilon_k+\varepsilon_s)^2+\Delta^2]^{1/2}-\varepsilon_{ext}\right\}}+1}-\frac{1}{e^{\beta\left\{[(\varepsilon_k+\varepsilon_s)^2+\Delta^2]^{1/2}+\varepsilon_{ext}\right\}}+1}\right) \qquad (22)$$

This expression is the same as B7 in reference [8].

The critical temperature may be calculated with $\Delta = 0$ in (22),

$$\frac{1}{NV} = \int_{-1}^{1}\frac{dx}{2}\int_{-\hbar\omega_D}^{\hbar\omega_D}\frac{d\varepsilon_k}{2|\varepsilon_k+\varepsilon_s|}\left(1-\frac{1}{e^{\frac{|\varepsilon_k+\varepsilon_s|-\varepsilon_{ext}}{kT_C}}+1}-\frac{1}{e^{\frac{|\varepsilon_k+\varepsilon_s|+\varepsilon_{ext}}{kT_C}}+1}\right) \qquad (23)$$

For Nb with $\Delta_0/kT_c(0) = 1.85$, $T_c(0) = 9.25$ K and $\xi_0 = 40$ nm [9], with $T_c(0)$ being the critical temperature with $v_s = 0$ and $\xi_0$ being the coherence length with $v_s = 0$ at 0 K temperature, the numerical solution of $\Delta$ and $T_c$ at 2K are plotted in Fig. 1. The calculation stopped at the critical velocity, which is determined by the critical current, where Cooper pairs start to break.

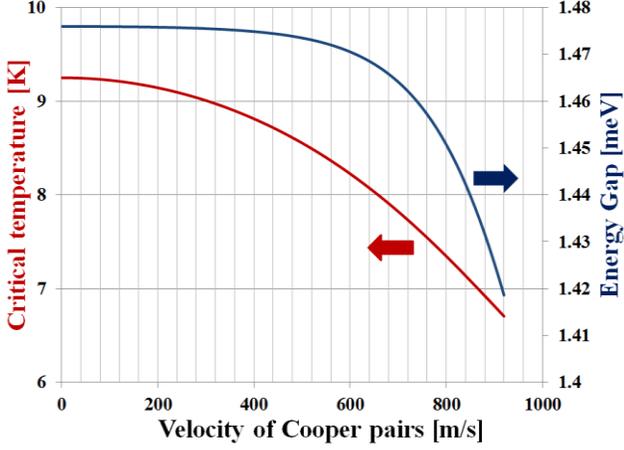

FIG. 1. Energy gap at 2 K and critical temperature versus Cooper pair velocity with parameters set for Nb. The energy gap at zero Cooper pair velocity is 1.476 meV.

## Derivation of BCS surface impedance

In order to calculate the SRF BCS surface impedance, one may start with the matrix elements of a single-particle scattering operator as in references [1, 2], using the new particle distribution equations (12)-(19) above.

Here we list the matrix elements of the first type of transition that corresponding to single occupancy of $k$ in the initial and of $k'$ in the final state in Table I. The other types shown in reference [1] can be listed in a similar way. Using the first row of Table I as an example, in the initial state, $k+q \uparrow$ is occupied, noted as X, $-k+q \downarrow$, $k'+q \uparrow$ and $-k'+q \downarrow$ are unoccupied, noted as 0. In the final state, only $k'+q \uparrow$ is occupied. Pair occupancy of ($k'+q \uparrow, -k'+q \downarrow$) in initial state and of ($k+q \uparrow, -k+q \downarrow$) in final state may be either ground (+) or excited (-), listed in the second column. For $k+q \uparrow$ single occupancy in initial state with both states ground, the initial energy is $E_{k+q \uparrow}$. For $k'+q \uparrow$ single occupancy in final state with both states ground, the final energy is $E_{k'+q \uparrow}$. The energy difference between them is listed in column 3. The probability of initial state ($k+q \uparrow, -k+q \downarrow$) to be X0 is $s_{k+q \uparrow}$ and the probability of initial state ($k'+q \uparrow, -k'+q \downarrow$) to be 00 is $(1-s_{-k'+q \downarrow}-s_{k'+q \uparrow}-p')$, the combination of these two probabilities is listed in column 4. Column 5 lists the matrix elements that are the same as Table II in reference [1]. In this column, $h$ and $h'$ are used to simplify the expression of $h_k$ and $h_{k'}$, respectively. One may refer to references [1, 2] for more detail about this table.

TABLE I. Matrix elements of single particle scattering operator with moving Cooper pairs.

| Wave functions | | | | Ground(+) or excited(-) | | Energy difference $W_i-W_f$ | Probability of initial state | Matrix elements | |
|---|---|---|---|---|---|---|---|---|---|
| Initial, $\psi_i$ | | Final, $\psi_f$ | | | | | | $c_{k'\uparrow}^* c_{k\uparrow}$ or $c_{-k\downarrow}^* c_{k\uparrow}$ | $c_{-k\downarrow}^* c_{-k'\downarrow}$ or $-c_{-k\downarrow}^* c_{k'\uparrow}$ |
| k | k' | k | k' | k | k' | | | | |
| X0 | 00 | 00 | X0 | + | + | $E_{k+q\uparrow}-E_{k'+q\uparrow}$ | $s_{k+q\uparrow}(1-s_{k'+q\downarrow}-s_{k'+q\uparrow}-p')$ | $[(1-h)(1-h')]^{1/2}$ | $-(hh')^{1/2}$ |
| X0 | XX | XX | X0 | - | - | $-E_{-k+q\downarrow}+E_{-k'+q\downarrow}$ | $s_{k+q\uparrow}p'$ | $(hh')^{1/2}$ | $-[(1-h)(1-h')]^{1/2}$ |
| | | | | + | - | $E_{k+q\uparrow}+E_{-k'+q\downarrow}$ | $s_{k+q\uparrow}p'$ | $-[(1-h)h']^{1/2}$ | $-[h(1-h')]^{1/2}$ |
| | | | | - | + | $-E_{-k+q\downarrow}-E_{k'+q\uparrow}$ | $s_{k+q\uparrow}(1-s_{k'+q\downarrow}-s_{k'+q\uparrow}-p')$ | $-[h(1-h')]^{1/2}$ | $-[(1-h)h']^{1/2}$ |
| X0 | 00 | 00 | 0X | + | + | $E_{k+q\uparrow}-E_{-k'+q\downarrow}$ | $s_{k+q\uparrow}(1-s_{k'+q\downarrow}-s_{k'+q\uparrow}-p')$ | $[(1-h)(1-h')]^{1/2}$ | $-(hh')^{1/2}$ |
| X0 | XX | XX | 0X | - | - | $-E_{-k+q\downarrow}+E_{k'+q\uparrow}$ | $s_{k+q\uparrow}p'$ | $(hh')^{1/2}$ | $-[(1-h)(1-h')]^{1/2}$ |
| | | | | + | - | $E_{k+q\uparrow}+E_{k'+q\uparrow}$ | $s_{k+q\uparrow}p'$ | $-[(1-h)h']^{1/2}$ | $-[h(1-h')]^{1/2}$ |
| | | | | - | + | $-E_{-k+q\downarrow}-E_{-k'+q\downarrow}$ | $s_{k+q\uparrow}(1-s_{k'+q\downarrow}-s_{k'+q\uparrow}-p')$ | $-[h(1-h')]^{1/2}$ | $-[(1-h)h']^{1/2}$ |
| 0X | 00 | 00 | X0 | + | + | $E_{-k+q\downarrow}-E_{k'+q\uparrow}$ | $s_{-k+q\downarrow}(1-s_{k'+q\downarrow}-s_{k'+q\uparrow}-p')$ | $[(1-h)(1-h')]^{1/2}$ | $-(hh')^{1/2}$ |
| 0X | XX | XX | X0 | - | - | $-E_{k+q\uparrow}+E_{-k'+q\downarrow}$ | $s_{-k+q\downarrow}p'$ | $(hh')^{1/2}$ | $-[(1-h)(1-h')]^{1/2}$ |
| | | | | + | - | $E_{-k+q\downarrow}+E_{-k'+q\downarrow}$ | $s_{-k+q\downarrow}p'$ | $-[(1-h)h']^{1/2}$ | $-[h(1-h')]^{1/2}$ |
| | | | | - | + | $-E_{k+q\uparrow}-E_{k'+q\uparrow}$ | $s_{-k+q\downarrow}(1-s_{k'+q\downarrow}-s_{k'+q\uparrow}-p')$ | $-[h(1-h')]^{1/2}$ | $-[(1-h)h']^{1/2}$ |
| 0X | 00 | 00 | 0X | + | + | $E_{-k+q\downarrow}-E_{-k'+q\downarrow}$ | $s_{-k+q\downarrow}(1-s_{k'+q\downarrow}-s_{k'+q\uparrow}-p')$ | $[(1-h)(1-h')]^{1/2}$ | $-(hh')^{1/2}$ |
| 0X | XX | XX | 0X | - | - | $-E_{k+q\uparrow}+E_{k'+q\uparrow}$ | $s_{-k+q\downarrow}p'$ | $(hh')^{1/2}$ | $-[(1-h)(1-h')]^{1/2}$ |
| | | | | + | - | $E_{-k+q\downarrow}+E_{k'+q\uparrow}$ | $s_{-k+q\downarrow}p'$ | $-[(1-h)h']^{1/2}$ | $-[h(1-h')]^{1/2}$ |
| | | | | - | + | $-E_{k+q\uparrow}-E_{-k'+q\downarrow}$ | $s_{-k+q\downarrow}(1-s_{k'+q\downarrow}-s_{k'+q\uparrow}-p')$ | $-[h(1-h')]^{1/2}$ | $-[(1-h)h']^{1/2}$ |

While under RF field with angular frequency $\omega$, the photon energy $\hbar(\omega-is)$ should be inserted into either the initial or the final state in Table I. Here a small positive parameter $s$, which will be set equal to zero in the final expression, has been introduced to obtain the real and imaginary part of surface impedance [2].

Based on the above analysis, the single-particle scattering operator, shown as equation (3.2) in reference [2] may be rewritten as,

$$L(\omega,\varepsilon,\varepsilon',x,x') = \frac{1}{4}\left(1+\frac{(\varepsilon+\varepsilon_s)(\varepsilon'+\varepsilon_s)+\Delta^2}{E_k E_{k'}}\right)\left(\frac{1}{E_{k+q\uparrow}-E_{k'+q\uparrow}+\hbar(\omega-is)}+\frac{1}{E_{k+q\uparrow}-E_{k'+q\uparrow}-\hbar(\omega-is)}\right)(f_{k'+q\uparrow}-f_{k+q\uparrow})+\frac{1}{4}\left(1-\frac{(\varepsilon+\varepsilon_s)(\varepsilon'+\varepsilon_s)+\Delta^2}{E_k E_{k'}}\right)\left(\frac{1}{E_{k+q\uparrow}+E_{k'+q\uparrow}+\hbar(\omega-is)}+\frac{1}{E_{k+q\uparrow}+E_{k'+q\uparrow}-\hbar(\omega-is)}\right)(1-f_{k'+q\uparrow}-f_{k+q\uparrow})$$
(24)

with $x$ (or $x'$) being cosine of the angle between $v_k$ (or $v_{k'}$) and $v_s$, consider the integration from -1 to 1 for $x$ and $x'$, equation (2) now becomes,

$$K(p) = \frac{-3}{4\pi\hbar v_F \lambda_L^2(0)}\int_0^\infty \frac{4}{(pR)^2}\left[\frac{\sin(pR)}{pR}-\cos(pR)\right]e^{-\frac{R}{l}}\times\frac{1}{4}\int_{-1}^1\int_{-1}^1 I(\omega,R,T,x,x')dxdx'\,dR \quad (25)$$

with $l$ the mean free path and

$$I(\omega,R,T,x,x') = \int_{-\infty}^\infty\int_{-\infty}^\infty\left[L(\omega,\varepsilon,\varepsilon',x,x')+\frac{1-2f(\varepsilon)}{\varepsilon'-\varepsilon}\right]\times\cos[\alpha(\varepsilon-\varepsilon')]d\varepsilon'\,d\varepsilon \quad (26)$$

with $\alpha = \frac{R}{\hbar p_F}$.

Expression (24) can be rearranged as,

$$L(\omega,\varepsilon,\varepsilon',x,x') = -\frac{1}{2}(1 - f(E_{k+q\uparrow}) - f(E_{-k+q\downarrow}))\left\{\frac{E_{k+q\uparrow}+\hbar(\omega-is)-\varepsilon_{ext}' + \frac{(\varepsilon+\varepsilon_S)(\varepsilon'+\varepsilon_S)+\Delta^2}{E}}{E_k'^2 - [E_{k+q\uparrow}-\varepsilon_{ext}' + \hbar(\omega-is)]^2} + \right.$$

$$\left.\frac{E_{k+q\uparrow}-\hbar(\omega-is)-\varepsilon_{ext}' + \frac{(\varepsilon+\varepsilon_S)(\varepsilon'+\varepsilon_S)+\Delta^2}{E}}{E_k'^2 - [E_{k+q\uparrow}-\varepsilon_{ext}' - \hbar(\omega-is)]^2}\right\} \quad (27)$$

And thus expression (26) in the limit $s\to 0$ is,

$$I(\omega,R,T,x,x') = -\pi i \int_{\Delta-\varepsilon_{ext}'-\hbar\omega}^{\infty}[1 - f(E_2 + \varepsilon_{ext}) - f(E_2 - \varepsilon_{ext})]\{g\cos[\alpha(\varepsilon_2 + \varepsilon_s)] - i\sin[\alpha(\varepsilon_2 + \varepsilon_s)]\}e^{i\alpha(\varepsilon_1+\varepsilon_s)}dE + \pi i \int_{\Delta-\varepsilon_{ext}'}^{\infty}[1 - f(E_1 + \varepsilon_{ext}) - f(E_1 - \varepsilon_{ext})]\{g\cos[\alpha(\varepsilon_1 + \varepsilon_s)] + i\sin[\alpha(\varepsilon_1 + \varepsilon_s)]\}e^{-i\alpha(\varepsilon_2+\varepsilon_s)}dE \quad (28)$$

with $E_1 = E + \varepsilon_{ext}$, $E_2 = E + \varepsilon_{ext}' + \hbar\omega$

$$\varepsilon_1 = \sqrt{E_1^2 - \Delta^2} - \varepsilon_s, \quad \varepsilon_2 = \sqrt{E_2^2 - \Delta^2} - \varepsilon_s$$

$$g = \frac{E_1 E_2 + \Delta^2}{(\varepsilon_1+\varepsilon_s)(\varepsilon_2+\varepsilon_s)} \text{ and } \alpha = \frac{R}{\hbar v_F}.$$

When the Cooper pair net velocity $v_s \to 0$, (28) reduces to equation (3.5) in reference [2].

One should notice that $\varepsilon_1$, $\varepsilon_2$ and $g$ are not always real in (28). The following expressions have been used to separate equation (28) into its real and imaginary parts:

$$Re[K(p)] = \frac{3}{4r\hbar v_F \lambda_L^2(0)} \int_{-1}^{1}\int_{-1}^{1} M(p,x,x')dxdx'$$

$$Im[K(p)] = \frac{3}{4r\hbar v_F \lambda_L^2(0)} \int_{-1}^{1}\int_{-1}^{1} N(p,x,x')dxdx' \quad (29)$$

with

$$M(p,x,x') = \frac{1}{2}\int_{\Delta-\varepsilon_{ext}}^{\infty}\{[f(E_2+\varepsilon_{ext})+f(E_2-\varepsilon_{ext})-f(E_1+\varepsilon_{ext})-f(E_1-\varepsilon_{ext})](g+1)S(a_-,a_3) - [2-f(E_2+\varepsilon_{ext})-f(E_2-\varepsilon_{ext})-f(E_1+\varepsilon_{ext})-f(E_1-\varepsilon_{ext})](g-1)S(a_+,a_3)\}dE + \int_{\Delta-\varepsilon_{ext}'-\hbar\omega}^{\Delta-\varepsilon_{ext}}[1-f(E_2+\varepsilon_{ext})-f(E_2-\varepsilon_{ext})][g_{1b}R(a_2,a_3+a_{1b})+S(a_2,a_3+a_{1b})]dE$$

$$N(p,x,x') = -\frac{1}{2}\int_{\Delta-\varepsilon_{ext}}^{\infty}[f(E_2+\varepsilon_{ext})+f(E_2-\varepsilon_{ext})-f(E_1+\varepsilon_{ext})-f(E_1-\varepsilon_{ext})][(g-1)R(a_+,a_3)+(g+1)R(a_-,a_3)]dE$$

while $\varepsilon_{ext} < \varepsilon'_{ext} + \hbar\omega$ and,

$$M(p,x,x') = \frac{1}{2}\int_{\Delta-\varepsilon_{ext}'-\hbar\omega}^{\infty}\{[f(E_2+\varepsilon_{ext})+f(E_2-\varepsilon_{ext})-f(E_1+\varepsilon_{ext})-f(E_1-\varepsilon_{ext})](g+1)S(a_-,a_3) - [2-f(E_2+\varepsilon_{ext})-f(E_2-\varepsilon_{ext})-f(E_1+\varepsilon_{ext})-f(E_1-\varepsilon_{ext})](g-1)S(a_+,a_3)\}dE + \int_{\Delta-\varepsilon_{ext}}^{\Delta-\varepsilon_{ext}'-\hbar\omega}[1-f(E_1+\varepsilon_{ext})-f(E_1-\varepsilon_{ext})][g_{2b}R(a_1,a_3+a_{2b})+S(a_1,a_3+a_{2b})]dE$$

$$N(p,x,x') = -\frac{1}{2}\int_{\Delta-\varepsilon_{ext}'-\hbar\omega}^{\infty}[f(E_2+\varepsilon_{ext})+f(E_2-\varepsilon_{ext})-f(E_1+\varepsilon_{ext})-f(E_1-\varepsilon_{ext})][(g-1)R(a_+,a_3)+(g+1)R(a_-,a_3)]dE \quad (30)$$

while $\varepsilon_{ext} > \varepsilon'_{ext} + \hbar\omega$,

and $\varepsilon_{1b} = \sqrt{\Delta^2 - E_1^2} - \varepsilon_s$, $\varepsilon_{2b} = \sqrt{\Delta^2 - E_2^2} - \varepsilon_s$

$$g_{1b} = \frac{E_1 E_2 + \Delta^2}{(\varepsilon_{1b}+\varepsilon_s)(\varepsilon_2+\varepsilon_s)}, \quad g_{2b} = \frac{E_1 E_2 + \Delta^2}{(\varepsilon_1+\varepsilon_s)(\varepsilon_{2b}+\varepsilon_s)}$$

$$a_1 = \frac{\varepsilon_1+\varepsilon_s}{p\hbar v_F}, \quad a_2 = \frac{\varepsilon_2+\varepsilon_s}{p\hbar v_F}, \quad a_3 = \frac{1}{pl}, \quad a_{ext} = \frac{\varepsilon_{ext}-\varepsilon_{ext}'}{p\hbar v_F},$$

$$a_{1b} = \frac{\varepsilon_{1b}+\varepsilon_s}{p\hbar v_F}, \quad a_{2b} = \frac{\varepsilon_{2b}+\varepsilon_s}{p\hbar v_F}, \quad a_+ = a_1 + a_2, \quad a_- = a_1 - a_2$$

$$R(a,b) = -\frac{b}{2} + \frac{ab}{4}\ln\frac{b^2+(1+a)^2}{b^2+(1-a)^2} + \frac{1+b^2-a^2}{4}\arctan\frac{2b}{b^2+a^2-1}$$

$$S(a,b) = \frac{a}{2} - \frac{ab}{2} arctan\frac{2b}{b^2+a^2-1} + \frac{1+b^2-a^2}{8} ln\frac{b^2+(1+a)^2}{b^2+(1-a)^2}$$

Here, all the elements in $M(p,x,x')$ and $N(p,x,x')$ are real. One now obtains here an analytical expression for the surface impedance by incorporating (29) and (30) into (3). The quadruple integral must be solved numerically.

## Surface impedance solution

The above analysis is valid for any conventional superconductor described by BCS theory. A Mathematica[TM] program has been developed to accomplish the calculation of this challenging integral. Using the following characteristic parameters: $\Delta_0/kT_c(0) = 1.85$, $T_c(0) = 9.25$ K, $\xi_0 = 40$ nm, $\lambda_L(0) = 32$ nm, and mean free path $\iota = 40$ nm [9], we calculate the surface impedance of niobium at 1.5 GHz and 2 K, with the results shown in Figure 2. The surface resistance and reactance values of the outermost layer of the superconductor are plotted as a function of the Cooper pair velocity in that layer. Since the supercurrent density varies both with depth into the surface and time, the surface impedance does as well.

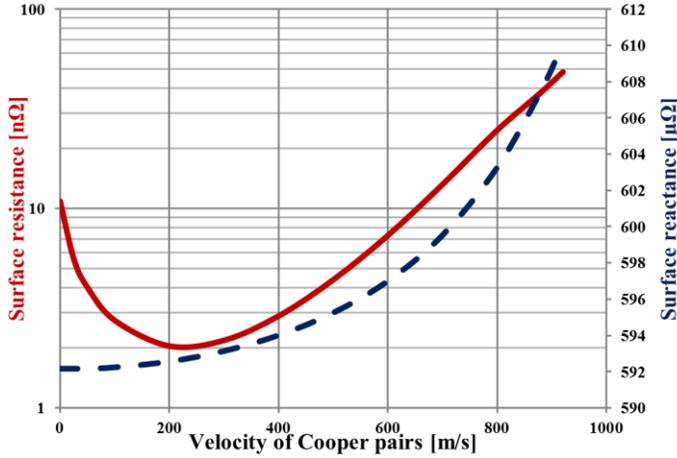

FIG. 2. Surface resistance (red line) and reactance (blue dashed line) versus Cooper pair velocity for Nb at 2 K and 1.5 GHz.

The surface resistance $R_s$, with a value of 10.9 n$\Omega$ at 0 m/s cooper pair velocity $v_s$, first decreases with increasing $v_s$, then increases with increasing $v_s$, with a minimum $R_s$ of 2.04 n$\Omega$ at 200 m/s $v_s$.

## Discussion

Qualitative insight into the change in $R_s$ may be obtained by considering the extreme anomalous limit approximation with $\alpha \to 0$ [2, 3]. In this approximation, equation (28) becomes,

$$K(p) = \frac{3}{4q\hbar v_F \lambda_L^2(0)} \frac{1}{4} \int_{-1}^{1}\int_{-1}^{1}\left\{-\pi i \int_{\Delta-\varepsilon_{ext}'-\hbar\omega}^{\infty}[1-f(E_2+\varepsilon_{ext})-f(E_2-\varepsilon_{ext})]gdE + \pi i \int_{\Delta-\varepsilon_{ext}}^{\infty}[1-f(E_1+\varepsilon_{ext})-f(E_1-\varepsilon_{ext})]gdE\right\}dxdx' \quad (31)$$

Based on (31), the energy loss is determined by the integration over energy and angles of the expression $[f(E_2+\varepsilon_{ext})+f(E_2-\varepsilon_{ext})- f(E_1+\varepsilon_{ext})-f(E_1-\varepsilon_{ext})]g$, the smaller the integration result, the lower the $R_s$. To give a physical explanation of the surface resistance changes with velocity of Cooper pairs, this integration is simplified in the low temperature approximation with $E >> kT$. In this approximation the excited particle distribution function simplifies to $f(E) = 1/e^{\beta E}$. The above expression then may be rewritten as $[f(E_2)-f(E_1)][f(\varepsilon_{ext})+f(-\varepsilon_{ext})]g$. In Mattis-Bardeen theory, which is in the weak field limit, scattering only happens between $E$ and $E+\hbar\omega$. In this extended theory, scattering occurs between energy levels $E_1$ and $E_2$, shown in the first term $[f(E_2)-f(E_1)]$ and is zero at $\varepsilon_{ext}=\varepsilon_{ext}'+\hbar\omega$. The contribution from $\varepsilon_{ext}<\varepsilon_{ext}'+\hbar\omega$ is partially cancelled by the contribution from $\varepsilon_{ext} > \varepsilon_{ext}' + \hbar\omega$, which leads to a smaller integration result. The low but non-zero field $R_s$ reduction is the result of this cancellation effect induced by the non-symmetric (or directional) net Cooper pair velocity. The second term, *[f($\varepsilon_{ext}$)+f(-$\varepsilon_{ext}$)]*, which appears as equation (3) in reference [7], combined with the reduction of energy gap with increasing $v_s$, yields an increase in $R_s$ with increasing $v_s$. With higher values of $v_s$, such as >200 m/s in the example of Figure 2, this increasing contribution more than compensates for the former effect.

The above analysis has been similarly applied to the RF surface impedance of superconductors in a static magnetic field [10,11]. The low-field $R_s$ reduction described here is attributed to the angle between the Cooper pair velocity driven by the RF field and the Fermi velocity, instead of the angle between the DC and RF fields as in Ref. [10].

## Calculation of effective surface resistance

For correspondence with operating resonant structures, one must obtain an appropriate average over space and time to calculate the effective surface resistance. In the purest sense, this is quite difficult because non-local and non-equilibrium effects should be considered. Such a treatment is beyond our present scope. We, rather, take the approximation that the distribution of the supercurrent is unaffected by the distribution of the quasiparticle current, which may be affected by the variation of the local surface resistance. In this case, with RF field $H(x,t) = H_0 sin(\omega t)e^{-x/\lambda}$, the Cooper pairs net velocity over depth and rf cycle is $v(x,t) = v_s cos(\omega t)e^{-x/\lambda}$. Where, as in [7],

$$v_S(H_0) = \frac{\pi}{2^{3/2}} \frac{H_0}{H_c} \frac{\Delta_0}{p_F} \tag{32}$$

and $H_c$ is the critical RF field.

We obtain the effective surface resistance by using a function derived from equation (3), such as is displayed in Figure 2 as the $R(v(x,t))$ term in the integration:

$$R = \frac{2}{H_0^2} \frac{\omega}{\pi} \int_0^{\pi/\omega} \frac{2}{\lambda} \int_0^\infty H(x,t)^2 R(v(x,t)) dx\, dt = \frac{2}{\pi} \int_0^\pi \int_0^1 R(v_s \cos\tau \sqrt{y}) \sin^2\tau\, dy\, d\tau \tag{33}$$

This average yet neglects the variation of surface field amplitude by location in a resonant structure such as characteristic of particular normal modes. For present purposes we take the approximation that modal field geometry is such that $H_{surf}$ amplitude is effectively either $H_0$ or 0. Then (33) can be useful. Such an approximation is not badly inappropriate for resonant niobium structures crafted for efficient acceleration of relativistic electrons [5].

This effective field-dependent BCS surface resistance for niobium as calculated using the $R_s$ function in Figure 2 is shown in Figure 3. This result shows encouraging coincidence with the recently reported effective surface resistance measurement between $H_{pk}$ of 5 mT and 90 mT on a CEBAF shape single cell cavity made from ingot niobium with 3 hours 1400°C high temperature baking by Dhakal *et al.* [12] after subtracting 1.8 nΩ temperature independent residual resistance. The corresponding quality factor versus peak magnetic flux density and effective electric accelerating gradient (as applies to particle accelerator applications) of the above measurements are plotted in Figure 4, and are compared with calculated results with 1.8 nΩ residual resistance added.

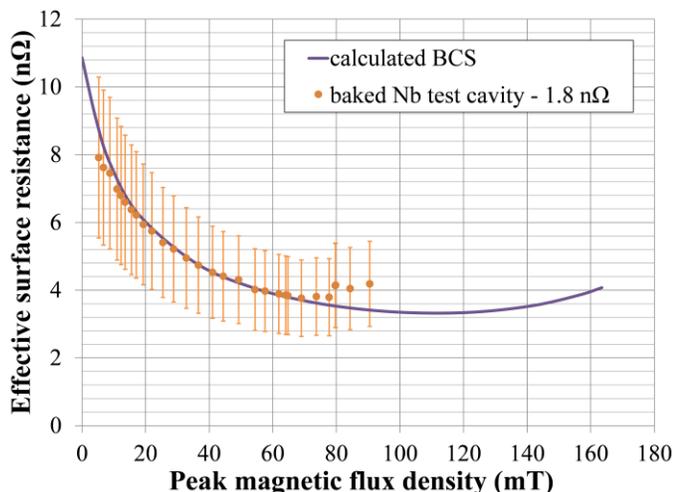

FIG. 3. Effective BCS surface resistance versus peak magnetic flux density for Nb at 2 K and 1.5 GHz. – calculated, ● measured single cell CEBAF cavity with 3 hour 1400°C bake .

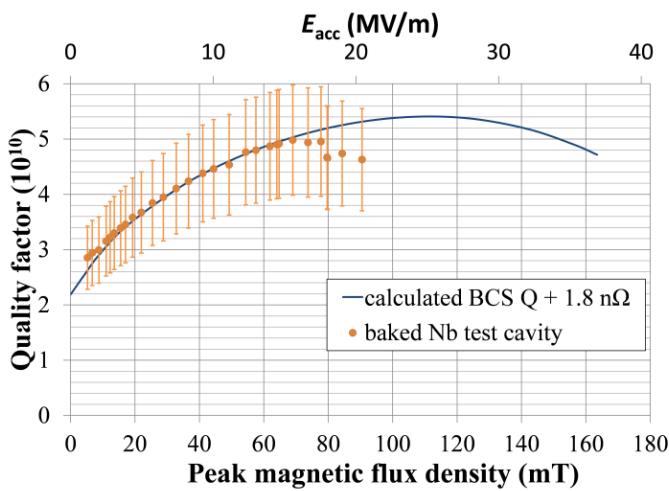

FIG. 4. Cavity quality factor versus peak magnetic flux density and corresponding accelerating gradient for Nb at 2 K and 1.5 GHz. – calculated field-dependent BCS Q with 1.8 nΩ additional residual resistance, ● single cell CEBAF test cavity with 3 hour 1400°C bake.

The decrease in effective surface resistance which produces the positive low-field $Q$ slope that is commonly observed to ~10 mT magnetic flux density is observed to continue to 80 mT in the baked cavity [12] and predicted by the analysis here to continue to ~110 mT.

## Summary

The electron states distribution at 0K and the probability of electron occupation with finite temperature have been calculated based on BCS theory and considering Cooper pairs to be moving coherently. These new expressions have been applied to Nb, the material of greatest interest to SRF applications, to calculate the numerical value of its effective surface impedance as a function of the peak surface rf magnetic flux density. Calculation of this non-linear BCS surface impedance indicates a minimum at intermediate field value, suggesting the prospect for lower cryogenic losses in a field regime of great interest to particle accelerator applications. With a set of typical niobium superconducting material parameters, the numerical calculation and approximate resulting effective surface impedance show encouraging coincidence with a recent measurement on a single cell cavity at Jefferson Lab with 3 hours 1400 °C bake and no subsequent wet chemistry. These results suggest that perhaps the rather desirable

10non-linear behavior observed reflects the absence of additional loss mechanisms and may reflect rather pure BCS mechanisms. Further exploration of the parameter space with a view to minimizing $R_s$ for niobium, as well as other superconductors, at maximum surface fields and various rf frequencies will continue and be reported soon.

## Acknowledgments

Authored by Jefferson Science Associates, LLC under U.S. DOE Contract No. DE-AC05-06OR23177. The U.S. Government retains a non-exclusive, paid-up, irrevocable, world-wide license to publish or reproduce this manuscript for U.S. Government purposes. The authors acknowledge the helpful discussions on this work with G. Ciovati, A. Gurevich, and F. He.

* reece@jlab.org